\documentclass{article}
\usepackage[T1]{fontenc} %
\usepackage[utf8]{inputenc} %
\usepackage{ismir,amsmath,cite,url}
\usepackage{amssymb}
\usepackage{graphicx}
\usepackage{textcomp}
\usepackage[neverdecrease]{paralist}
\usepackage{multirow}
\usepackage{booktabs}
\usepackage{fmtcount}
\usepackage{siunitx}
\usepackage{soul}
\usepackage{xcolor}
\usepackage{listings}
\usepackage{subcaption}

\newif\ifarxiv
\arxivtrue

\title{Lyrics Transcription for Humans: A Readability-Aware Benchmark}

\multauthor{%
Ondřej Cífka \hspace{0.5cm} Hendrik Schreiber \hspace{0.5cm} Luke Miner \hspace{0.5cm} Fabian-Robert Stöter \vspace{0.1cm}}{%
AudioShake\\
{\tt\small ondrej@audioshake.ai}, {\tt\small fabian@audioshake.ai}
}

\def\authorname{O.\ Cífka, H.\ Schreiber, L.\ Miner, and F.-R.\ Stöter}

\usepackage[bookmarks=false,pdfauthor={\authorname},pdfsubject={\papersubject},hidelinks]{hyperref}
\usepackage[capitalize,nameinlink]{cleveref}

\newcommand{\punct}{\texttt{P}} %
\newcommand{\paren}{\texttt{B}} %
\newcommand{\nl}{\texttt{L}} %
\newcommand{\sect}{\texttt{S}}

\newcommand{\m}[1]{{\small\texttt{#1}}}  %

\makeatletter
\newcommand*{\textoverline}[1]{$\overline{\hbox{#1}}\m@th$}
\makeatother

\newcommand{\casewer}{{WER$'$}}

\newcommand{\ctr}[1]{\multicolumn{1}{@{}c@{}}{#1}}
\newcommand{\nan}{\multicolumn{1}{@{}c@{}}{---}}

\definecolor{op-hit}{HTML}{DDDDDD}
\definecolor{op-case}{HTML}{AAAAAA}
\definecolor{op-near}{HTML}{FFEE00}
\definecolor{op-sub}{HTML}{FF7F0D}
\definecolor{op-ins}{HTML}{1DB1FF}
\definecolor{op-del}{HTML}{FF4444}

\newcommand{\cul}[2]{{\setulcolor{#1}\setul{-0.05ex}{0.4ex}\ul{\itshape #2}}}

\newcommand{\AudioShake}{AudioShake} %

\begin{document}

\maketitle
\begin{abstract}
Writing down lyrics for human consumption involves not only accurately capturing word sequences, but also incorporating punctuation and formatting for clarity and to convey contextual information.
This includes song structure, emotional emphasis, and contrast between lead and background vocals. 
While automatic lyrics transcription (ALT) systems have advanced beyond producing unstructured strings of words and are able to draw on wider context, ALT benchmarks have not kept pace and continue to focus exclusively on words.
To address this gap, we introduce Jam-ALT, a comprehensive lyrics transcription benchmark. %
The benchmark features a complete revision of the JamendoLyrics dataset, in adherence to industry standards for lyrics transcription and formatting, along with
evaluation metrics designed to capture and assess the lyric-specific nuances, laying the foundation for improving the readability of lyrics.
We apply the benchmark to recent transcription systems and present additional error analysis, as well as an experimental comparison with a classical music dataset.
\end{abstract}
\section{Introduction}\label{sec:introduction}
Recent general-purpose automatic speech recognition (ASR) models trained on large datasets \cite{baevski-2020-wav2vec2,pmlr-v202-radford23a} have shown a remarkable level of generalization, even improving the performance of automatic lyrics transcription (ALT) \cite{ou-2022-wav2vec2-alt,zhuo-2023-lyricwhiz,WangLLSJ23}.
Remarkably, these state-of-the-art ASR models are able to take in larger temporal contexts and produce natural text with long-term coherence which, in the case of Whisper \cite{pmlr-v202-radford23a}, includes punctuation and capitalization \cite{gris2023}.
One may therefore ask how well these capabilities transfer from speech to lyrics.
Moreover, producing a high-quality lyrics transcript suitable for user-facing music industry applications (e.g.\ to be displayed on streaming platforms or lyrics websites) presents some unique challenges, namely the need for specific formatting (e.g.\ line break placement, parentheses around background vocals) \cite{apple-lyrics-guidelines,lyricfind-guidelines,musixmatch-lyric-guidelines}.
This calls for a new approach to ALT evaluation and development that accounts for these distinctive nuances.

In ASR, the primary goal is a clear representation of what was said. To that end, formatting is helpful for improving the readability of transcripts~\cite{Jones03_readability}.
Likewise, fillers like \emph{um}, \emph{uh}, \emph{like}, and \emph{you know} can be omitted to improve readability. 
Recent work~\cite{hwer} attempts to formalize this concern for clarity, proposing a novel metric geared towards assessing human readability.
It employs human labelers, instructed to %
disregard filler words while, on the other hand, taking account of punctuation and capitalization errors that impact readability or alter the meaning of the text.

In music, on the other hand, lyrics are not simply a means of communicating meaning; they are a form of artistic expression, closely tied to the rhythm, melody, and emotionality of the song.
For this reason, lyrics transcription requires a different set of considerations.
Line breaks, often missing or arbitrarily placed in speech transcripts, are essential in lyrics for capturing rhyme, meter, and musical phrasing.
Fillers like \emph{oh yeah}, non-word sounds like \emph{la-la-la} and contractions such as \emph{I'ma} (vs.\ \emph{I'm gonna}, \emph{I am going to}) have prosodic significance, and their omission would disrupt the song’s rhythm and rhyme scheme.
Far from being an impediment to readability, they are key to any faithful rendition of a song for artist and fan alike.

\begin{figure}[t]
    \centering
    \includegraphics[width=\linewidth,trim={0.2cm 0.2cm 0.2cm 0.2cm},clip]{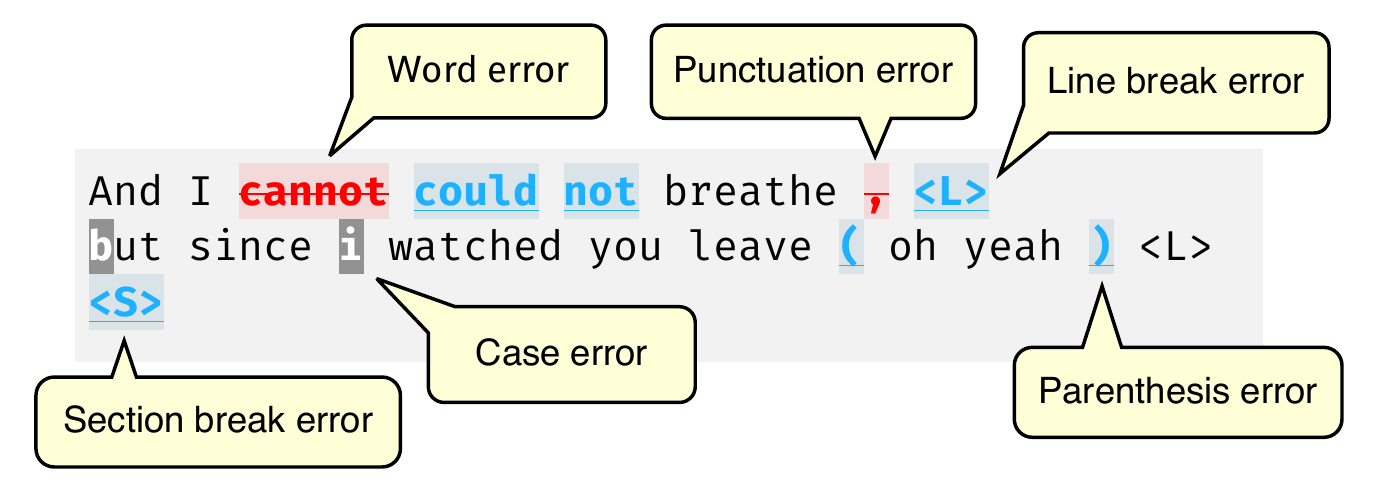}
    \caption{Error types captured by our metrics. Each token is classified as a word, punctuation mark, or parenthesis (enclosing background vocals). Special tokens are added in place of line and section breaks. Each token type is covered by a separate metric; differences in letter case are handled separately.}
    \label{fig:teaser}
\end{figure}

We believe that readability-aware models for lyrics transcription have the potential to facilitate novel applications extending beyond the realms of metadata extraction and relatively crude karaoke subtitles.
However, in order to advance in this research direction, the ability to accurately evaluate ALT systems in the aforementioned aspects is vital.
To the best of our knowledge, existing ALT literature not only overlooks readability, but evaluates on datasets (e.g.\ \cite{hsu2010,meseguer_brocal_2018,stoller2019end,wang2022}) that have not been designed specifically for ALT and lack some or all of the desirable features discussed above.

One of the datasets widely adopted by recent works \cite{gupta2020,demirel2021mstrenet,demirel2021low,ou-2022-wav2vec2-alt,zhuo-2023-lyricwhiz} as an ALT test set is JamendoLyrics \cite{stoller2019end}, originally a lyrics alignment benchmark.
Its most recent (``MultiLang'') version \cite{durand-2023-contrastive} contains four languages and a diverse set of genres, making it attractive as a testbed for lyrics-related tasks.
However, we found that, in addition to lacking in the aspects discussed above, the lyrics are sometimes inaccurate or incomplete.
While such lyrics may be perfectly acceptable as input for lyrics alignment (and indeed representative of a real-world scenario for that task), they are less suitable as a target for ALT.

To address these issues and help to guide future ALT research, we present 
the \textbf{Jam-ALT} benchmark,
consisting of:
\begin{inparaenum}[(1)]
\item a revised version of JamendoLyrics MultiLang following a newly created annotation guide that unifies the music industry’s conventions for lyrics transcription and formatting (in particular, regarding punctuation, line breaks, letter case, and non-word vocal sounds);
\item a comprehensive set of automated evaluation metrics designed to capture and distinguish different types of errors relevant to (1).
\end{inparaenum}
The dataset and the implementation of the metrics are available via the project website.\footnote{\url{https://audioshake.github.io/jam-alt/}}
Additionally, to explore the applicability of the proposed metrics to other datasets, we present results on the \emph{Schubert Winterreise Dataset} (SWD) \cite{WeissZAMKVG21}.

\section{Dataset}
\label{sec:dataset}
Our first contribution is a revision of the JamendoLyrics MultiLang dataset \cite{durand-2023-contrastive} to make it more suitable as a lyrics transcription test set.
Different sets of guidelines for lyrics transcription and formatting exist within the music industry; we consider guidelines by Apple \cite{apple-lyrics-guidelines}, LyricFind \cite{lyricfind-guidelines}, and Musixmatch \cite{musixmatch-lyric-guidelines}, from which we extracted the following general rules:
\begin{compactenum}
    \item Only transcribe words and vocal sounds audible in the recording; exclude credits, section labels, style markings, non-vocal sounds, etc.
    \item Break lyrics up into lines and sections; separate sections by a single blank line.
    \item Include each word, line and section as many times as heard. Do not use shorthands to indicate repetitions.
    \item Start each line with a capital letter; respect standard capitalization rules for each language.
    \item Respect standard punctuation rules, but never end a line with a comma or a period.
    \item Use standard spelling, including stan\-dard\-ized spell\-ing for slang where appropriate.
    \item Mark elisions (incomplete words) and contractions with an apostrophe.
    \item Transcribe background vocals and non-word vocal sounds if they contribute to the content of the song.
    \item Place background vocals in parentheses.
\end{compactenum}
The original JamendoLyrics dataset adheres to rules 1, 3, and 8, partially 2 and 6 (up to some missing diacritics, misspellings, and misplaced line breaks), but lacks punctuation and is lowercase, thus ignoring rules 4, 5, 7, and 9. Moreover, as mentioned above, we found that the lyrics do not always accurately correspond to the audio.

To address these issues, we revised the lyrics in order for them to obey all of the above rules and to match the recordings as closely as possible.
As the above rules are fairly unspecific, we created a detailed annotation guide where we have attempted to resolve minor discrepancies among the source guidelines \cite{apple-lyrics-guidelines,lyricfind-guidelines,musixmatch-lyric-guidelines} and fill in missing details (including language-specific nuances).
This annotation guide is released together with the dataset.

Each lyric file was revised by a single annotator proficient in the language, then reviewed by two other annotators. %
In coordination with the authors of \cite{durand-2023-contrastive}, one of the 20 French songs was removed following the detection of potentially harmful content.

Examples of lyrics before and after revision can be found on the project website. %

\section{Metrics}
\label{sec:metrics}
In this section, we first discuss our adaptation of the conventional \emph{word error rate} (WER) metric and then our proposed precision and recall measures for punctuation and formatting.
Our goal here is to design a comprehensive set of metrics that covers all possible transcription errors while allowing us to distinguish between different types of errors (see \cref{fig:teaser} for a visual overview of the error types). Note, however, that our goal is \emph{not} to create metrics that completely align with the rules put forth in \cref{sec:dataset} or correlate with a specific notion of readability; the metrics should be general enough to apply to any plain-text lyrics dataset and adapt to its formatting style.

\subsection{Word Error Rates}
\label{sec:wer}
The standard speech recognition metric, WER, is defined as the edit distance (a.k.a.\ Levenshtein distance) between the \emph{hypothesis} (predicted transcription) and the \emph{reference} (ground-truth transcript), normalized by the length of the reference.
If $D$, $I$, and $S$ are the number of word \emph{deletions}, \emph{insertions}, and \emph{substitutions} respectively, for the minimal sequence of edits needed to turn the reference into the hypothesis, and $H$ is the number of unchanged words (\emph{hits}), then:
\begin{equation}
    \text{WER} = \frac{S+D+I}{S+D+H} = \frac{S+D+I}{N},
    \label{eq:wer}
\end{equation}
where $N$ is the total number of reference words.

Typically, the hypothesis and the reference are pre-processed to make the metric insensitive to variations in punctuation, letter case, and whitespace, but no single standard pre-processing procedure exists.
In this work, we apply Moses-style \cite{koehn-etal-2007-moses} punctuation normalization and tokenization, then remove all non-word tokens.
Before computing the WER, %
we lowercase each token to make the metric case-insensitive, but also keep track of the token's original form.
To then measure the error in letter case, 
for every \emph{hit} in the minimal edit sequence, we compare the original forms of the hypothesis and the reference token and count an error if they differ. %
We then compute a \emph{case-sensitive word error rate} \casewer\ as:
\begin{equation}
    \text{\casewer} = \frac{S+D+I+E_\text{case}}{S+D+H} = \text{WER}+\frac{E_\text{case}}{N},
    \label{eq:caswewer}
\end{equation}
where $E_\text{case}$ is the number of casing errors.
We include both variants \eqref{eq:wer} and \eqref{eq:caswewer} in our benchmark.

\subsection{Punctuation and Line Breaks}
\label{sec:pr-metrics}
Since the output of ASR systems traditionally lacks punctuation, a common ASR post-pro\-cess\-ing step~-- \emph{punctuation restoration} \cite{pais-2022-capitalization}~-- consists of recovering it.
This task is usually evaluated using precision and recall:
\begin{equation}
    \begin{gathered}
    P = \frac{\text{\small\# correctly predicted symbols}}{\text{\small\# predicted symbols}},\\
    R = \frac{\text{\small\# correctly predicted symbols}}{\text{\small\# expected symbols}}.
    \end{gathered}
    \label{eq:pr}
\end{equation}
In this original setting where the system only inserts punctuation and the words remain intact, computing the metrics is trivial.
In contrast, in our end-to-end setting, the hypothesis and the reference may use different words, and hence computing the numerator in \cref{eq:pr} requires an alignment between the two.
We leverage the same alignment as used in \cref{sec:wer}, but computed on text that includes punctuation.
Moreover, we extend this approach to account for line breaks, which, though traditionally ignored in speech data, are particularly important for lyrics.

We use the pre-processing from \cref{sec:wer}, but preserve punctuation tokens and, as in \cite{matusov-etal-2019-customizing,karakanta-etal-2020-42}, add special tokens in place of line and section breaks; this leaves us with five token types: word \texttt{W}, punctuation \punct, parenthesis \paren\ (separate due to its distinctive function), line break \nl, and section break \sect.%
\footnote{We define a section break as one or more blank lines. Hence, every section break is explicitly preceded by a line break in our representation.}
After computing the alignment between the hypothesis tokens and the reference tokens, we iterate through it in order to count, for each token type $T\in\{\texttt{W},\punct,\paren,\nl,\sect\}$, its number of deletions $D_T$, insertions $I_T$, substitutions $S_T$, and hits $H_T$.
In general, each edit operation is simply attributed to the type of the token affected (e.g.\ the insertion of a punctuation mark counts towards $I_\punct$).
However, a substitution of a token of type $T$ by a token of type $T'\neq T$ is counted as two operations: a deletion of type $T$ (counting towards $D_T$) and an insertion of type $T'$ (counting towards $I_{T'}$).

We can now use these counts to define a precision, recall, and F-1 metric for each token type:
\begin{equation}
    \begin{gathered}
    P_T = \frac{H_T}{H_T+S_T+I_T}, \hskip0.5em\relax R_T = \frac{H_T}{H_T+S_T+D_T},\\
    F_T = \frac{2}{P_T^{-1}+R_T^{-1}}.
    \end{gathered}
\end{equation}

\section{Results}
\label{sec:results}

\subsection{Benchmark Results} %
\begin{figure}
    \centering
    \includegraphics[width=0.9\linewidth,trim={0.2cm 0.2cm 0.2cm 0.2cm},clip]{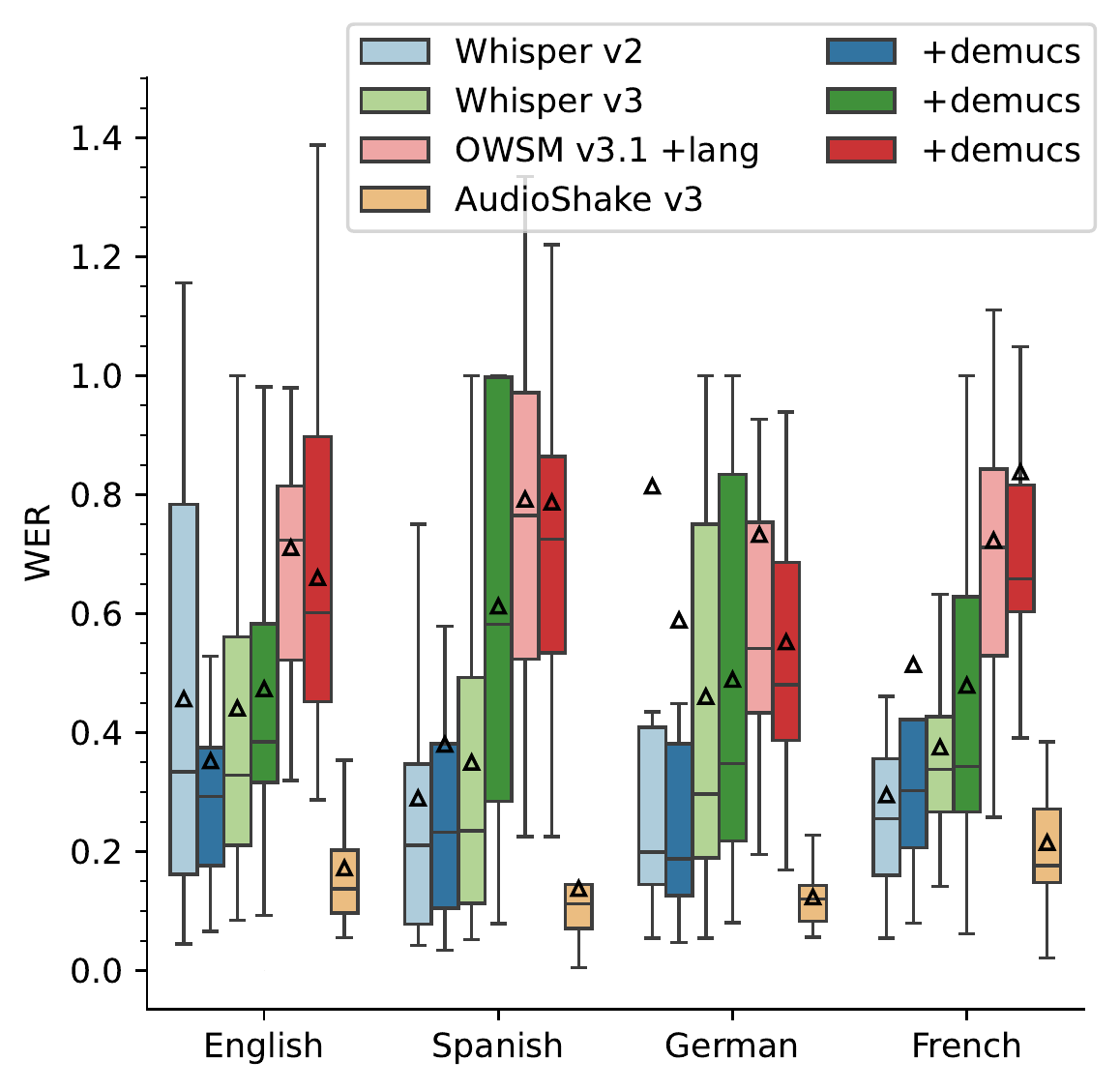}
    \caption{Song-level word error rates by language. Note that strong outliers occur; for clarity, they are not displayed here, but affect the means, which are indicated by triangles.}
    \label{fig:wer-boxplot}
\end{figure}

\cref{tab:results} shows the performance of various transcription systems on our benchmark. \cref{fig:wer-boxplot} shows the distributions of song-level word error rates by language.

\begin{table*}%
    \centering%
    \scalebox{0.88}{%
    \renewcommand{\arraystretch}{0.9}%
    \setlength\aboverulesep{0.35ex}%
    \setlength\belowrulesep{0.65ex}%
    \begin{tabular}{@{~}l@{~~}r@{~~}r@{~~}r@{~~}r@{~~}r@{~~}rr@{~~}r@{~~}r@{~~}r@{~~}r@{~~}rr@{~~}rr@{~~}rr@{~~}r@{~}}
    \toprule
     & \multicolumn{6}{c}{All languages} & \multicolumn{6}{c}{English} & \multicolumn{2}{c}{Spanish} & \multicolumn{2}{c}{German} & \multicolumn{2}{c}{French} \\
     \cmidrule(r){2-7} \cmidrule(rl){8-13} \cmidrule(rl){14-15} \cmidrule(rl){16-17} \cmidrule(l){18-19}
     & \small WER & \small\casewer & $F_\punct$ & $F_\paren$ & $F_\nl$ & $F_\sect$ & \small WER & \small\casewer & $F_\punct$ & $F_\paren$ & $F_\nl$ & $F_\sect$ & \small WER & \small\casewer & \small WER & \small\casewer & \small WER & \small\casewer \\
    \midrule
Whisper v2 & 37.8 & 42.1 & 44.2 & \nan & 69.3 & 3.3 & 43.8 & 47.5 & 31.5 & \nan & 63.0 & 11.2 & 25.8 & 31.5 & 54.5 & 59.3 & 27.7 & 31.1 \\
\quad\m{+lang} & \bfseries 27.9 & \bfseries 32.6 & \bfseries 45.0 & \nan & 70.4 & \bfseries 3.7 & 39.7 & 43.7 & 34.9 & \nan & 65.5 & \bfseries 11.6 & \bfseries 21.9 & \bfseries 27.7 & \bfseries 19.9 & \bfseries 26.0 & \bfseries 27.1 & \bfseries 30.5 \\
\quad\m{+demucs} & 44.5 & 49.8 & 41.6 & \nan & 61.2 & \nan & \bfseries 33.3 & \bfseries 39.1 & \bfseries 42.2 & \nan & 53.9 & \nan & 39.6 & 46.5 & 65.2 & 70.4 & 43.3 & 46.9 \\
\qquad\m{+lang} & 33.5 & 39.3 & 39.4 & \nan & 60.6 & \nan & 35.6 & 41.3 & 41.8 & \nan & 53.4 & \nan & 34.9 & 42.2 & 23.9 & 30.4 & 38.2 & 42.1 \\
Whisper v3 & 35.5 & 39.7 & 43.0 & \nan & 73.5 & 1.0 & 37.7 & 42.5 & 41.4 & \nan & 71.5 & 2.6 & 28.6 & 33.6 & 40.7 & 44.6 & 34.7 & 38.0 \\
\quad\m{+lang} & 32.6 & 37.2 & 43.7 & \nan & \bfseries 73.9 & 0.6 & 36.4 & 41.4 & 41.8 & \nan & \bfseries 72.5 & 2.6 & 22.4 & 28.0 & 35.9 & 40.4 & 34.7 & 38.0 \\
\quad\m{+demucs} & 48.0 & 51.6 & 33.0 & \nan & 65.7 & \nan & 43.0 & 47.2 & 25.8 & \nan & 66.9 & \nan & 61.5 & 64.9 & 43.5 & 47.4 & 44.9 & 48.2 \\
\qquad\m{+lang} & 46.6 & 50.4 & 33.7 & \nan & 65.8 & \nan & 43.0 & 47.2 & 25.8 & \nan & 66.9 & \nan & 58.6 & 62.1 & 40.8 & 44.9 & 44.9 & 48.3 \\
OWSM v3.1\m{+lang} & 69.3 & 75.0 & 22.5 & \bfseries 0.6 & 37.8 & \nan & 68.6 & 74.0 & 22.3 & \nan & 42.7 & \nan & 73.3 & 78.5 & 63.3 & 71.8 & 71.6 & 75.7 \\
\quad\m{+demucs} & 66.5 & 72.6 & 20.0 & 0.0 & 41.1 & \nan & 63.4 & 69.4 & 21.5 & \bfseries 0.0 & 47.3 & \nan & 70.8 & 76.0 & 51.8 & 62.0 & 78.5 & 82.1 \\
\midrule
LyricWhiz & \nan & \nan & \nan & \nan & \nan & \nan & 24.6 & 28.0 & 34.0 & \nan & 74.0 & 1.4 & \nan & \nan & \nan & \nan & \nan & \nan \\
\AudioShake\ v3 & 16.1 & 20.1 & 57.0 & 29.4 & 84.4 & 73.9 & 17.3 & 20.9 & 65.3 & 37.9 & 84.3 & 84.8 & 12.6 & 17.7 & 12.6 & 17.5 & 20.8 & 23.5 \\
    \midrule[\heavyrulewidth]
    JamendoLyrics & 11.1 & 29.6 & \nan & \nan & 93.3 & 85.3 & 14.4 & 29.6 & \nan & \nan & 88.1 & 77.9 & 14.0 & 29.1 & 5.0 & 37.6 & 10.3 & 23.3 \\
    \bottomrule
    \end{tabular}%
    }
    \caption{Benchmark results (all metrics shown as percentages). WER is word error rate, %
    \casewer\ is case-sensitive WER, the rest are F-measures. \m{+demucs} indicates vocal separation using HTDemucs; \m{+lang} indicates that the language of each song was provided to the model instead of relying on auto-detection. Whisper results are averages over 5 runs with different random seeds, LyricWhiz over 2 runs; OWSM and \AudioShake\ are deterministic, hence the results are from a single run. The best results achieved by open-source systems are shown in {\bfseries bold}. LyricWhiz and \AudioShake\ are listed separately, because they rely on proprietary technology. The last row shows metrics computed between the original JamendoLyrics dataset as the hypotheses and our revision as the reference.
    For full results by language, see \ifarxiv\cref{tab:full-results} in the appendix.\else the project website.\fi}
    \label{tab:results}
\end{table*}

\begin{table}
\centering
\scalebox{0.88}{%
\renewcommand{\arraystretch}{0.9}
\setlength\aboverulesep{0.35ex}
\setlength\belowrulesep{0.65ex}
\begin{tabular}{@{~}l@{~~}r@{~~}r@{~~}rr@{~~}r@{~~}r@{~~}r@{~}}
\toprule
& \multicolumn{3}{c}{All} & \ctr{EN} & \ctr{ES} & \ctr{DE} & \ctr{FR} \\
\cmidrule(r){2-4} \cmidrule(l){5-8}
& \small WER & $F_\nl$ & $F_\sect$ & \multicolumn{4}{c}{\small WER} \\
\midrule
Whisper v2 & 39.1 & 70.0 & \bfseries 2.8 & 43.0 & 31.7 & 54.7 & 28.0 \\
\quad\m{+lang} & \bfseries 28.8 & 71.0 & 2.6 & 38.8 & \bfseries 27.9 & \bfseries 19.8 & \bfseries 27.4 \\
\quad\m{+demucs} & 46.2 & 61.5 & \nan & \bfseries 33.6 & 43.9 & 65.5 & 44.1 \\
\qquad\m{+lang} & 34.8 & 61.2 & \nan & 36.1 & 39.3 & 23.9 & 38.9 \\
Whisper v3 & 37.7 & 71.6 & 1.0 & 39.3 & 34.5 & 40.8 & 36.1 \\
\quad\m{+lang} & 34.9 & \bfseries 72.3 & 0.6 & 38.0 & 28.9 & 36.0 & 36.1 \\
\quad\m{+demucs} & 49.6 & 65.3 & \nan & 44.3 & 65.8 & 43.5 & 45.7 \\
\qquad\m{+lang} & 48.3 & 65.4 & \nan & 44.3 & 63.1 & 40.8 & 45.7 \\
OWSM v3.1\m{+lang} & 70.3 & 39.0 & \nan & 69.9 & 75.7 & 63.5 & 71.9 \\
\quad\m{+demucs} & 67.5 & 41.6 & \nan & 65.0 & 72.7 & 51.7 & 79.1 \\
\midrule
LyricWhiz & \nan & \nan & \nan & 23.7 & \nan & \nan & \nan \\
\AudioShake\ v3 & 19.4 & 82.3 & 64.5 & 22.5 & 18.7 & 13.8 & 21.7 \\
\midrule[\heavyrulewidth]
Jam-ALT & 11.5 & 94.0 & 85.1 & 15.7 & 14.4 & 5.0 & 10.4 \\
\bottomrule
\end{tabular}%
}
\caption{Results with the original JamendoLyrics (i.e.\ before revision) as reference.
The last row corresponds to our revision.
See also the caption of \cref{tab:results}.}
\label{tab:jamendolyrics-results}
\end{table}

We include two recent, freely available models capable of transcribing long, unsegmented audio: Whisper~\cite{pmlr-v202-radford23a} (\texttt{large-v2} and \texttt{large-v3}) and OWSM 3.1~\cite{owsm31} (\texttt{owsm\_v3.1\_ebf}).
For both models, we use Whisper-style long-form transcription with a beam size of 5.
Both models have language identification capabilities, but may perform better if the correct language is specified; for Whisper, we evaluate both options, while for OWSM, for simplicity, we only evaluate with the language provided.
For Whisper, which exhibits great variation between runs due to its stochastic decoding strategy, we report averages over 5 runs.
We optionally use HTDemucs \cite{rouard-2023-htdemucs} to isolate the vocals from the input audio.%

Whisper and OWSM are general-purpose speech recognition models and are not designed for lyrics transcription.
To make a fairer comparison, we apply simple post-processing to their outputs to improve the formatting:
(1)~The models do not produce line breaks, but split their output into timestamped segments; we insert line breaks between these segments.
(2)~We remove unwanted end-of-line punctuation (all non-word characters except for \texttt{!?\textquotesingle\textquotedbl»)}) and uppercase the first letter of every line.\footnote{Although we observed that this transformation tends to improve the outputs for Whisper and OWSM, in general, it may make evaluation results worse if the line break predictions are incorrect. For this reason, we do not include this step as a fixed part of our benchmark.}

We also evaluate LyricWhiz \cite{zhuo-2023-lyricwhiz}, a lyrics transcription system combining Whisper with the commercially available instruction-following language model ChatGPT \cite{chatgpt}.
We report averages over two outputs per song (English only), kindly provided by the LyricWhiz authors.
Finally, as an example of an ALT system built with formatting and readability in mind,
we include our in-house lyrics transcription system, which integrates vocal separation.

As a first general observation, consistent with previous studies \cite{zhuo-2023-lyricwhiz,WangLLSJ23}, the performance of Whisper models is relatively good, considering that they were not specifically designed for lyrics transcription.
Among the formatting metrics, we highlight a high accuracy in line break prediction.
This shows that, although the segments output by Whisper do not always impose a meaningful structure, in music, they do in many cases coincide with lyric lines.

Somewhat counter-intuitively, for Whisper, inputting isolated vocals (\m{+demucs}) tends to substantially degrade the results (with the single exception of \texttt{large-v2} for English).
Whisper's language identification mechanism also turns out to have a significant effect, in that disabling it and instead inputting the known language of the song (\m{+lang}) tends to result in a sizeable drop in WER, especially on languages different from English.
This suggests that the language detected by Whisper is often incorrect.%

We also observe that Whisper v3 does not necessarily perform better on lyrics than v2. In fact, the WER increases from \num{27.9} to \num{32.6} when comparing Whisper v2 \m{+lang} to v3 \m{+lang}.

The improvement of LyricWhiz over plain Whisper in terms of WER is clear and even sharper than reported in \cite{zhuo-2023-lyricwhiz}.
We also see some improvement in terms of line breaks and punctuation.

Regarding OWSM, its performance is far behind Whisper, with differences far larger than reported in \cite{owsm31} for speech, strongly suggesting that OWSM is poorly suited for ALT, at least without finetuning.
With isolated vocals as input, the error is slightly reduced, but still large.

As for our own system, it outperforms all of the above on all metrics shown in \cref{tab:results}, by a large margin, e.g.\ with a \SI{57}{\percent} reduction in overall WER compared to Whisper v2.
It is also the only one achieving acceptable accuracy for parentheses (\paren) and section breaks (\sect).

\subsection{Effect of Revisions}
The revisions described in \cref{sec:dataset} have enabled us to compute metrics related to letter case and punctuation, features that are missing from the original dataset.
However, the revisions also involved correcting words and line breaks; to measure the effect of these corrections, we present in \cref{tab:jamendolyrics-results} the relevant metrics computed on the original JamendoLyrics data.
Comparing \cref{tab:results,tab:jamendolyrics-results}, we note that the revisions have mostly improved the results, notably reducing the overall WER (by \num{1.7}, or \SI{5.3}{\percent}, on average) for all systems, with Spanish seeing the sharpest drop (\num{4.7}, or \SI{17.4}{\percent}, on average, likely due to frequently missing accents in the original data).
The general trends~-- in particular, the ranking based on WER and $F_\nl$~-- remain mostly unchanged.

To quantify the extent of our revisions more directly, we also evaluate both versions of the lyrics against each other and include the results as the last row in \cref{tab:results,tab:jamendolyrics-results}.
Remarkably, in terms of word tokens, Jam-ALT differs from JamendoLyrics by about \SI{11}{\percent} (around \SI{15}{\percent} for English and Spanish), which is substantially more than the difference between system performance on the two dataset versions.
One potential explanation is that a significant number of the corrections correspond to low-intelligibility singing, which is prone to transcription errors, or to background vocals, which are susceptible to being omitted by transcription systems.

\subsection{Error Analysis}
\label{sec:error-analysis}
In this section, we further analyze the errors made by selected systems on our benchmark.

\begin{figure}
    \centering
    \includegraphics[width=0.92\linewidth,trim={0.2cm 0.2cm 0.2cm 0.2cm},clip]{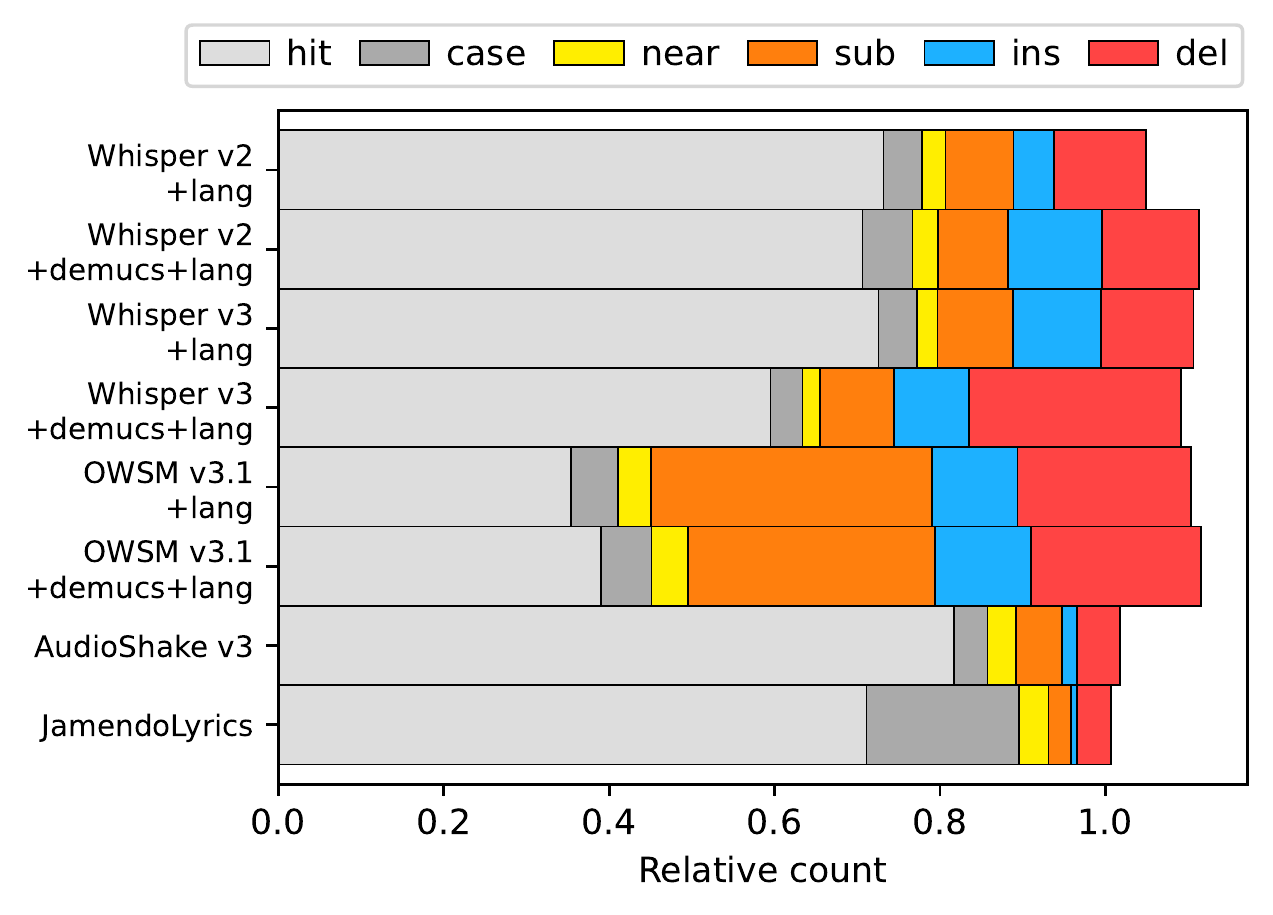}
    \caption[]{Word edit operation frequencies on our benchmark (one run per system). %
    \cul{op-near}{Near} are substitutions that differ in few characters, \cul{op-sub}{sub} are the remaining substitutions.
    \cul{op-case}{case} are hits with case errors, \cul{op-hit}{hit} are the remaining (case-sensitive) hits.
    The rest are \emph{\cul{op-ins}{ins}ertions} and \emph{\cul{op-del}{del}etions}.
    The frequencies are normalized by the reference length, so that:
    \begin{compactitem}
        \item $\text{\cul{op-hit}{hit}}+\text{\cul{op-case}{case}}+\text{\cul{op-near}{near}}+\text{\cul{op-sub}{sub}}+\text{\cul{op-del}{del}}=1$,
        \item $\text{WER}=\text{\cul{op-near}{near}}+\text{\cul{op-sub}{sub}}+\text{\cul{op-ins}{ins}}+\text{\cul{op-del}{del}}$,
        \item $\text{\casewer}-\text{WER}=\text{\cul{op-case}{case}}$,
        \item $\text{\cul{op-hit}{hit}}+\text{\cul{op-case}{case}}+\text{\cul{op-near}{near}}+\text{\cul{op-sub}{sub}}+\text{\cul{op-ins}{ins}}$ corresponds to the length of the prediction.
    \end{compactitem}}
    \label{fig:error-counts}
\end{figure}

First, we visualize in \cref{fig:error-counts} how each type of edit operation contributes to the WER.
Besides the basic edit operations (hits, substitutions, insertions, deletions), we include \emph{case errors} from \cref{sec:wer}; that is, a hit with a difference in letter case is shown as a case error instead.
Moreover, to account for small spelling differences, we consider a substitution as a \emph{near hit} when
the replacement differs from the reference in at most two letters.\footnote{
More precisely, we count a \emph{near hit} if, after removing apostrophes from the two words, their character-level Levenshtein distance is at most 2, and strictly less than half the length of the longer of the two words. Examples include \emph{an}/\emph{and}, \emph{gon'}/\emph{gonna}, \emph{there}/\emph{their}/\emph{they}/\emph{them}, but not \emph{a}/\emph{an} or \emph{this}/\emph{that}.
}

With Whisper, we observe that inputting separated vocals causes more insertions (and longer output) in v2, but more deletions (and shorter output) in v3.
Upon inspecting the outputs, we find that Whisper has a general tendency to omit parts of the lyrics (often the entire song) and instead produce generic or irrelevant text, and that this is more frequent with separated vocals, especially with v3.
On the other hand, OWSM shows a slight improvement with separated vocals, but its predictions contain significantly more substitutions, suggesting that they are more often incorrect on a word-by-word basis.

Next, we focus on errors in punctuation and formatting and investigate how often different token types are substituted for each other. %
To this end, we count the edit operations as in \cref{sec:pr-metrics}, but preserve the information about substitutions across the four non-word token types (\punct, \paren, \nl, \sect).
We then present this information in a form akin to a \emph{confusion matrix}, adding a special ``null'' token type $\varnothing$ to account for insertions and deletions.

\begin{figure*}
    \centering
    \subfloat[Whisper v2 \m{+lang}]{\includegraphics[trim={0.2cm 0.2cm 0.2cm 0.2cm},clip,scale=0.68]{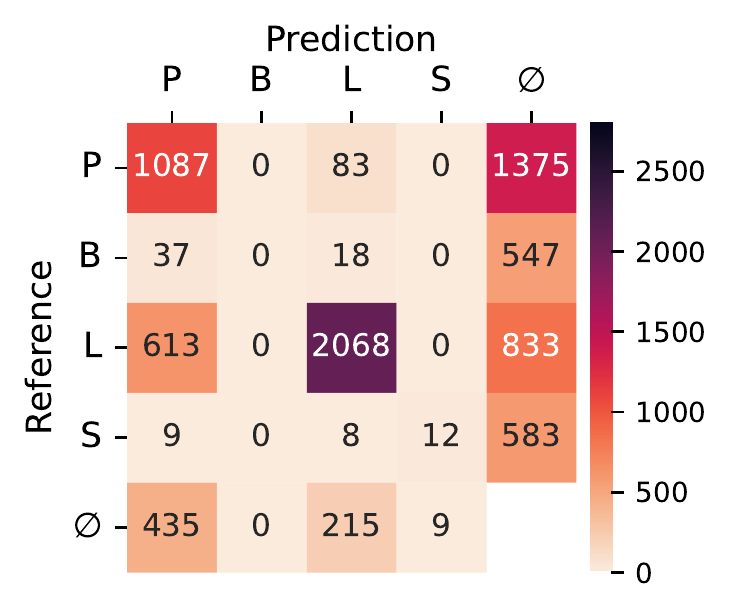}}\hspace{0.5cm}
    \subfloat[Whisper v2 \m{+demucs+lang}\label{fig:conf-mat-whisper-demucs}]{\includegraphics[trim={0.2cm 0.2cm 0.2cm 0.2cm},clip,scale=0.68]{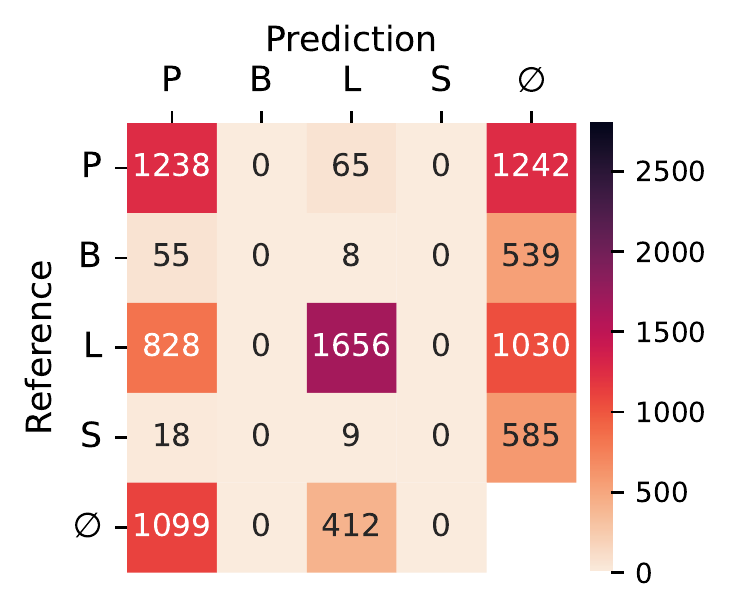}}\hspace{0.5cm}
    \subfloat[\AudioShake\ v3]{\includegraphics[trim={0.2cm 0.2cm 0.2cm 0.2cm},clip,scale=0.68]{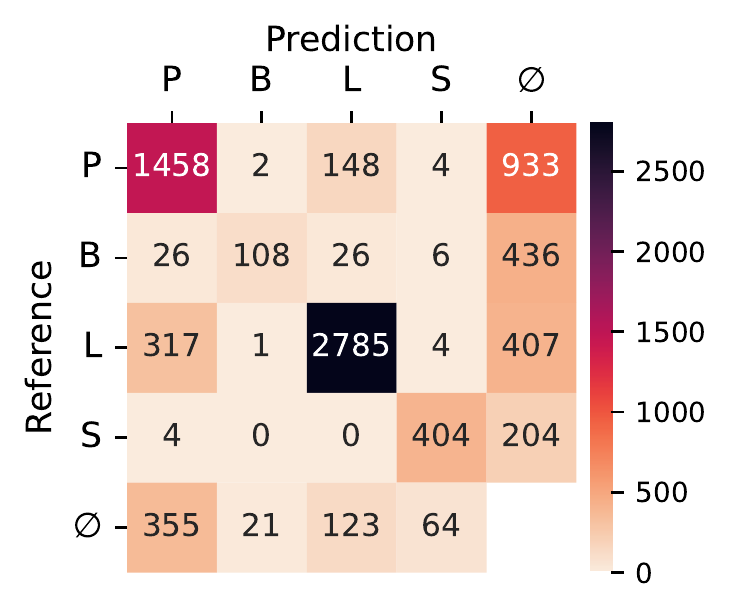}}
    \caption{Edit operation counts on non-word (punctuation and formatting) tokens by token type
    (\punct~= punctuation, \paren~= parenthesis, \nl~= line break, \sect~= section break).
    $\varnothing$ denotes the absence of a token, i.e.\ it stands for insertion (on the \emph{reference} axis) or deletion (on the \emph{prediction} axis).
    Substitution of/by a \emph{word} token is counted as an insertion/deletion, respectively.
    Only a single run per system is considered.}
    \label{fig:conf-mat}
\end{figure*}

The result is shown in \cref{fig:conf-mat} for three selected systems.
Most errors are insertions and deletions, but another frequent type of error is the replacement of a line break by a punctuation mark, especially in Whisper models.
This is explained by the fact that our guidelines forbid most end-of-line punctuation, and hence, when transcription omits a line break, inserting a punctuation mark in its place is often needed to maintain grammatical correctness.

By manual inspection of the transcriptions, we find that Whisper tends to produce much longer lines than in the reference and frequently outputs periods (forbidden by our annotation guide as a sentence separator) and, occasionally, spuriously repeated punctuation.

\subsection{Schubert Winterreise Dataset}
To explore the application of the proposed metrics to other datasets, we additionally perform an evaluation on the \emph{Schubert Winterreise Dataset} (SWD) \cite{WeissZAMKVG21}. SWD comprises nine audio versions of Franz Schubert’s 24-song cycle \emph{Winterreise}, along with symbolic representations, lyrics, and other annotations.
An example of Romantic music based on early \ordinalnum{19} century German poetry, it contrasts with JamendoLyrics and presents an interesting challenge for ALT.
For our evaluation, we pick a single version, \texttt{SC06} (a 2006 live recording of singer Randall Scarlata), one of the two with audio publicly available.

The lyrics in SWD are formatted as poems~-- containing line and section breaks~--, but their spelling and punctuation, mirroring an 1827 edition of the score \cite{SchubertWinterreiseScore}, does not exactly match our annotation guide.
To make them adhere to our punctuation and capitalization rules, we apply a simple transformation to the lyrics: replace all unwanted punctuation (\texttt{.;:-}) with commas, then remove all end-of-line commas and uppercase the first letter of each line.
Note, however, that even after this transformation, the lyrics' obsolete spelling~-- predating the 1996 German orthography reform~-- violates our annotation guide to some extent (mainly in the usage of the letter \emph{ß} and the treatment of elisions), which is expected to distort the WER.

\begin{figure}
    \centering
    \includegraphics[width=0.92\linewidth,trim={0.2cm 0.2cm 0.2cm 0.2cm},clip]{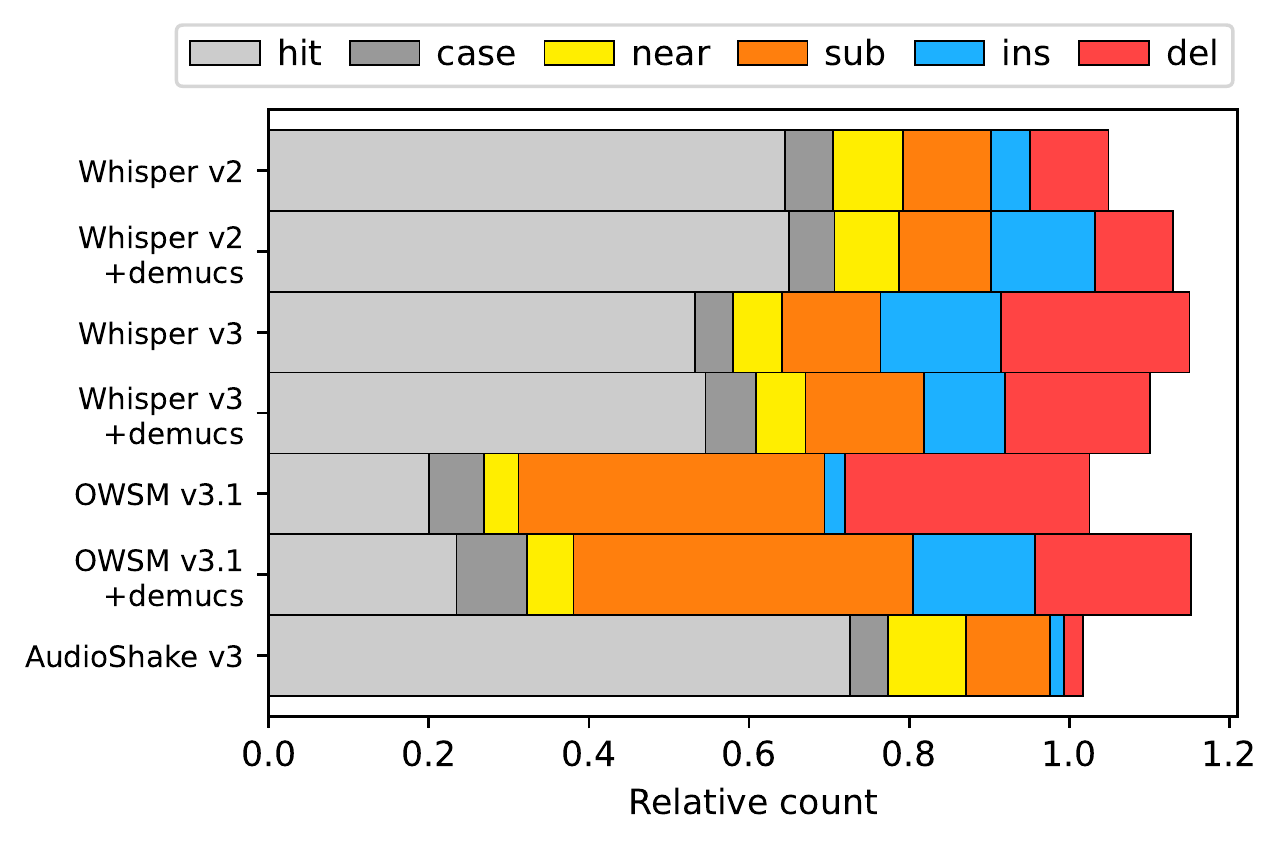}
    \caption{Word edit operation frequencies on SWD. See the caption of \cref{fig:error-counts}.}%
    \label{fig:swd-error-counts}
\end{figure}

We evaluate all models with the language provided (i.e.\ disabling language identification).
The results are shown in \cref{tab:swd-results} and further error analysis in \cref{fig:swd-error-counts}.
We notice substantially worse performance on SWD than the German section of our benchmark (\cref{tab:results}): for example, WER for Whisper v2 \m{+lang} increased from \num{19.9} to \num{34.5}.
This likely reflects the more challenging nature of the dataset, but also possibly the mismatched spelling, as suggested by a higher frequency of near hits (see \cref{fig:swd-error-counts}) than seen in \cref{sec:error-analysis} (\cref{fig:error-counts}).

\begin{table}
    \centering
    \scalebox{0.88}{%
    \renewcommand{\arraystretch}{0.9}
    \setlength\aboverulesep{0.35ex}
    \setlength\belowrulesep{0.65ex}
    \begin{tabular}{lrrrrr}
    \toprule
    & \ctr{\small WER} & \ctr{\small\casewer} & \ctr{$F_\punct$} & \ctr{$F_\nl$} & \ctr{$F_\sect$} \\
    \midrule
Whisper v2 & \bfseries 34.5 & \bfseries 40.4 & \bfseries 42.6 & \bfseries 66.2 & --- \\
\quad\m{+demucs} & 41.4 & 47.2 & 38.0 & 61.4 & --- \\
Whisper v3 & 59.0 & 63.8 & 40.0 & 63.6 & --- \\
\quad\m{+demucs} & 52.3 & 58.6 & 34.7 & 63.3 & 0.0 \\
OWSM v3.1 & 75.6 & 82.5 & 12.9 & 39.6 & \bfseries 4.9 \\
\quad\m{+demucs} & 82.9 & 91.8 & 17.0 & 39.2 & --- \\
    \midrule
\AudioShake\ v3 & 24.3 & 29.1 & 50.9 & 80.0 & 72.0 \\
    \bottomrule
    \end{tabular}
    }
    \caption{Results on performance \texttt{SC06} from SWD.
    Only punctuation (\texttt{P}), line breaks (\texttt{L}) and section breaks (\texttt{S}) are included, as the ground truth lyrics do not contain any parentheses.
    Whisper results are averages over 5 runs with different random seeds.
    The best result in each column, excluding \AudioShake, is shown in {\bfseries bold}. For full results, see \ifarxiv\cref{tab:full-swd-results} in the appendix.\else the project website.\fi}
    \label{tab:swd-results}
\end{table}

\section{Discussion}
\label{sec:discussion}
Given our focus on formatting and punctuation, the question arises to what extent they are in fact dependent on the audio.
In particular, could line and section boundaries be accurately predicted just from the textual context, e.g.\ based on metrical patterns, rhyme, syntax, and semantics?
To answer this, we suggest an experiment where a human annotator is tasked with formatting given lyrics first without and then with access to the audio.
Such a task would, however, be highly time-consuming and require expert annotators unfamiliar with the songs.
As a proxy, one might instead train a \emph{formatting restoration} model on lyrics or use a general-purpose instruction-following language model.
Our attempts in this regard have only had limited success and we therefore leave such experiments for future work.

Another issue is that there may not always be a single correct division into lines and sections. For example, in a song with relatively short lines, it may be acceptable to join pairs of adjacent lines, especially in the absence of rhyme. Likewise, 4-line sections may be joined to create 8-line sections and so forth. %
However, it is not obvious how to relax the metrics to allow for this kind of variation.
Doing so rigorously would likely require additional annotations, which is contrary to our goal of creating a set of generally applicable metrics.
A possible solution compatible with this idea is
to create multiple references and pick the best-scoring one during evaluation.

\section{Conclusion}
We have proposed Jam-ALT,
a new benchmark for ALT, based on the music industry's lyrics guidelines.
Our results show how existing systems differ in their performance on different aspects of the task, and we hope that the benchmark will be beneficial in guiding future ALT research.

\section{Acknowledgment}
We would like to thank Lau\-ra I\-bá\-ñez, Pa\-me\-la Ode, Ma\-thieu Fon\-taine, Clau\-dia Fal\-ler, Con\-sta\-nti\-nos Di\-mi\-triou, and Ka\-te\-ři\-na Apo\-lí\-no\-vá for their help with data annotation.
We are also thankful to Mei\-nard Mül\-ler and Hans-Ul\-rich Be\-ren\-des for their helpful comments on the manuscript.

\bibliography{bibliography}

\ifarxiv
\onecolumn
\appendix

\begin{table*}[ph]
\small
\renewcommand{\arraystretch}{0.9}
\setlength\heavyrulewidth{0.15ex}
\setlength\lightrulewidth{0.1ex}
\setlength\aboverulesep{0.3ex}
\setlength\belowrulesep{0.55ex}
\centering
\begin{tabular}{llr@{~}rr@{~}r@{~}rr@{~}r@{~}rr@{~}r@{~}rr@{~}r@{~}r}
\toprule
& & \multicolumn{2}{c}{Words} & \multicolumn{3}{c}{Punctuation} & \multicolumn{3}{c}{Parentheses} & \multicolumn{3}{c}{Line breaks} & \multicolumn{3}{c}{Section breaks} \\
\cmidrule(lr){3-4} \cmidrule(lr){5-7} \cmidrule(lr){8-10} \cmidrule(lr){11-13} \cmidrule(l){14-16}
Language & System & WER & \casewer & \ctr{$P_\punct$} & \ctr{$R_\punct$} & \ctr{$F_\punct$} & \ctr{$P_\paren$} & \ctr{$R_\paren$} & \ctr{$F_\paren$} & \ctr{$P_\nl$} & \ctr{$R_\nl$} & \ctr{$F_\nl$} & \ctr{$P_\sect$} & \ctr{$R_\sect$} & \ctr{$F_\sect$} \\
\midrule
\multirow[t]{12}{*}{All} & Whisper v2 & 37.8 & 42.1 & 48.3 & 40.7 & 44.2 & \nan & 0.0 & \nan & \bfseries 87.3 & 57.5 & 69.3 & \bfseries 55.2 & 1.7 & 3.3 \\
 & \quad\m{+lang} & \bfseries 27.9 & \bfseries 32.6 & 47.8 & 42.5 & \bfseries 45.0 & \nan & 0.0 & \nan & 86.6 & 59.3 & 70.4 & 53.3 & \bfseries 1.9 & \bfseries 3.7 \\
 & \quad\m{+demucs} & 44.5 & 49.8 & 38.1 & \bfseries 45.9 & 41.6 & \nan & 0.0 & \nan & 74.2 & 52.1 & 61.2 & \nan & 0.0 & \nan \\
 & \qquad\m{+lang} & 33.5 & 39.3 & 35.4 & 44.4 & 39.4 & \nan & 0.0 & \nan & 79.1 & 49.1 & 60.6 & \nan & 0.0 & \nan \\
 & Whisper v3 & 35.5 & 39.7 & \bfseries 50.4 & 37.5 & 43.0 & \nan & 0.0 & \nan & 76.9 & 70.4 & 73.5 & 37.5 & 0.5 & 1.0 \\
 & \quad\m{+lang} & 32.6 & 37.2 & 50.1 & 38.7 & 43.7 & \nan & 0.0 & \nan & 75.2 & \bfseries 72.6 & \bfseries 73.9 & 32.4 & 0.3 & 0.6 \\
 & \quad\m{+demucs} & 48.0 & 51.6 & 37.5 & 29.4 & 33.0 & \nan & 0.0 & \nan & 76.3 & 57.6 & 65.7 & \nan & 0.0 & \nan \\
 & \qquad\m{+lang} & 46.6 & 50.4 & 38.2 & 30.2 & 33.7 & \nan & 0.0 & \nan & 76.0 & 58.0 & 65.8 & \nan & 0.0 & \nan \\
 & OWSM v3.1\m{+lang} & 69.3 & 75.0 & 24.7 & 20.7 & 22.5 & \bfseries 4.3 & \bfseries 0.3 & \bfseries 0.6 & 80.7 & 24.6 & 37.8 & \nan & 0.0 & \nan \\
 & \quad\m{+demucs} & 66.5 & 72.6 & 19.7 & 20.3 & 20.0 & 0.0 & 0.0 & 0.0 & 83.4 & 27.3 & 41.1 & \nan & 0.0 & \nan \\
 \cmidrule{2-16}
 & \AudioShake\ v3 & 16.1 & 20.1 & 62.1 & 52.7 & 57.0 & 81.8 & 17.9 & 29.4 & 90.4 & 79.3 & 84.4 & 83.8 & 66.0 & 73.9 \\
 & JamendoLyrics & 11.1 & 29.6 & \nan & 0.0 & \nan & \nan & 0.0 & \nan & 96.2 & 90.7 & 93.3 & 84.6 & 85.9 & 85.3 \\
\midrule
\multirow[t]{13}{*}{English} & Whisper v2 & 43.8 & 47.5 & 41.3 & 25.5 & 31.5 & \nan & \bfseries 0.0 & \nan & 81.2 & 51.6 & 63.0 & 52.3 & 6.3 & 11.2 \\
 & \quad\m{+lang} & 39.7 & 43.7 & 42.4 & 29.8 & 34.9 & \nan & \bfseries 0.0 & \nan & 80.6 & 55.3 & 65.5 & \bfseries 53.0 & \bfseries 6.6 & \bfseries 11.6 \\
 & \quad\m{+demucs} & \bfseries 33.3 & \bfseries 39.1 & 41.4 & \bfseries 43.1 & \bfseries 42.2 & \nan & \bfseries 0.0 & \nan & 76.2 & 41.8 & 53.9 & \nan & 0.0 & \nan \\
 & \qquad\m{+lang} & 35.6 & 41.3 & 42.7 & 40.9 & 41.8 & \nan & \bfseries 0.0 & \nan & 75.7 & 41.2 & 53.4 & \nan & 0.0 & \nan \\
 & Whisper v3 & 37.7 & 42.5 & 48.0 & 36.4 & 41.4 & \nan & \bfseries 0.0 & \nan & 75.5 & 68.0 & 71.5 & 33.3 & 1.4 & 2.6 \\
 & \quad\m{+lang} & 36.4 & 41.4 & \bfseries 48.0 & 37.1 & 41.8 & \nan & \bfseries 0.0 & \nan & 74.8 & \bfseries 70.3 & \bfseries 72.5 & 33.3 & 1.4 & 2.6 \\
 & \quad\m{+demucs} & 43.0 & 47.2 & 32.5 & 21.5 & 25.8 & \nan & \bfseries 0.0 & \nan & 70.2 & 63.9 & 66.9 & \nan & 0.0 & \nan \\
 & \qquad\m{+lang} & 43.0 & 47.2 & 32.5 & 21.5 & 25.8 & \nan & \bfseries 0.0 & \nan & 70.2 & 63.9 & 66.9 & \nan & 0.0 & \nan \\
 & OWSM v3.1\m{+lang} & 68.6 & 74.0 & 22.9 & 21.7 & 22.3 & \nan & \bfseries 0.0 & \nan & 77.6 & 29.5 & 42.7 & \nan & 0.0 & \nan \\
 & \quad\m{+demucs} & 63.4 & 69.4 & 20.2 & 23.1 & 21.5 & \bfseries 0.0 & \bfseries 0.0 & \bfseries 0.0 & \bfseries 82.1 & 33.2 & 47.3 & \nan & 0.0 & \nan \\
 \cmidrule{2-16}
 & LyricWhiz & 24.6 & 28.0 & 49.0 & 26.2 & 34.0 & \nan & 0.0 & \nan & 87.5 & 64.1 & 74.0 & 100.0 & 0.3 & 1.4 \\
 & \AudioShake\ v3 & 17.3 & 20.9 & 68.0 & 62.8 & 65.3 & 81.7 & 24.6 & 37.9 & 88.3 & 80.7 & 84.3 & 87.0 & 82.8 & 84.8 \\
 & JamendoLyrics & 14.4 & 29.6 & \nan & 0.0 & \nan & \nan & 0.0 & \nan & 93.6 & 83.3 & 88.1 & 73.6 & 82.8 & 77.9 \\
\midrule
\multirow[t]{12}{*}{Spanish} & Whisper v2 & 25.8 & 31.5 & 54.2 & \bfseries 51.5 & \bfseries 52.8 & \nan & \bfseries 0.0 & \nan & \bfseries 86.2 & 61.4 & 71.7 & \bfseries 100.0 & 0.6 & \bfseries 3.1 \\
 & \quad\m{+lang} & \bfseries 21.9 & \bfseries 27.7 & 54.5 & 50.7 & 52.5 & \nan & \bfseries 0.0 & \nan & 85.4 & 61.5 & 71.5 & 51.8 & \bfseries 1.3 & 3.1 \\
 & \quad\m{+demucs} & 39.6 & 46.5 & 39.8 & 41.2 & 40.4 & \nan & \bfseries 0.0 & \nan & 77.1 & 44.7 & 56.6 & \nan & 0.0 & \nan \\
 & \qquad\m{+lang} & 34.9 & 42.2 & 32.2 & 36.8 & 34.3 & \nan & \bfseries 0.0 & \nan & 70.5 & 41.9 & 52.6 & \nan & 0.0 & \nan \\
 & Whisper v3 & 28.6 & 33.6 & 56.1 & 34.2 & 42.5 & \nan & \bfseries 0.0 & \nan & 75.1 & 72.4 & 73.7 & \nan & 0.0 & \nan \\
 & \quad\m{+lang} & 22.4 & 28.0 & \bfseries 57.3 & 36.3 & 44.5 & \nan & \bfseries 0.0 & \nan & 71.9 & \bfseries 77.3 & \bfseries 74.5 & 0.0 & 0.0 & 0.0 \\
 & \quad\m{+demucs} & 61.5 & 64.9 & 41.1 & 26.9 & 32.4 & \nan & \bfseries 0.0 & \nan & 80.1 & 38.8 & 52.3 & \nan & 0.0 & \nan \\
 & \qquad\m{+lang} & 58.6 & 62.1 & 42.0 & 29.3 & 34.4 & \nan & \bfseries 0.0 & \nan & 79.2 & 41.8 & 54.7 & \nan & 0.0 & \nan \\
 & OWSM v3.1\m{+lang} & 73.3 & 78.5 & 12.1 & 6.9 & 8.8 & \bfseries 0.0 & \bfseries 0.0 & \bfseries 0.0 & 80.6 & 18.6 & 30.2 & \nan & 0.0 & \nan \\
 & \quad\m{+demucs} & 70.8 & 76.0 & 14.5 & 6.5 & 9.0 & \nan & \bfseries 0.0 & \nan & 82.4 & 21.0 & 33.5 & \nan & 0.0 & \nan \\
 \cmidrule{2-16}
 & \AudioShake\ v3 & 12.6 & 17.7 & 71.9 & 46.8 & 56.7 & 25.0 & 2.3 & 4.2 & 84.6 & 78.7 & 81.5 & 76.0 & 59.0 & 66.4 \\
 & JamendoLyrics & 14.0 & 29.1 & \nan & 0.0 & \nan & \nan & 0.0 & \nan & 94.3 & 93.1 & 93.7 & 79.0 & 82.1 & 80.5 \\
\midrule
\multirow[t]{12}{*}{German} & Whisper v2 & 54.5 & 59.3 & 39.9 & 57.7 & 47.1 & \nan & \bfseries 0.0 & \nan & \bfseries 93.5 & 56.0 & 70.0 & \nan & 0.0 & \nan \\
 & \quad\m{+lang} & \bfseries 19.9 & \bfseries 26.0 & 39.2 & 63.1 & 48.4 & \nan & \bfseries 0.0 & \nan & 92.2 & 58.6 & 71.7 & \nan & 0.0 & \nan \\
 & \quad\m{+demucs} & 65.2 & 70.4 & 40.0 & 63.5 & 49.1 & \nan & \bfseries 0.0 & \nan & 66.2 & \bfseries 68.5 & 67.3 & \nan & 0.0 & \nan \\
 & \qquad\m{+lang} & 23.9 & 30.4 & 38.6 & \bfseries 67.6 & \bfseries 49.2 & \nan & \bfseries 0.0 & \nan & 84.9 & 60.5 & 70.6 & \nan & 0.0 & \nan \\
 & Whisper v3 & 40.7 & 44.6 & \bfseries 42.8 & 52.8 & 47.3 & \nan & \bfseries 0.0 & \nan & 79.1 & 64.5 & 71.1 & \bfseries 50.0 & \bfseries 0.6 & \bfseries 1.2 \\
 & \quad\m{+lang} & 35.9 & 40.4 & 41.5 & 55.3 & 47.4 & \nan & \bfseries 0.0 & \nan & 76.8 & 66.2 & 71.1 & \nan & 0.0 & \nan \\
 & \quad\m{+demucs} & 43.5 & 47.4 & 38.7 & 54.9 & 45.4 & \nan & \bfseries 0.0 & \nan & 84.0 & 62.9 & \bfseries 71.9 & \nan & 0.0 & \nan \\
 & \qquad\m{+lang} & 40.8 & 44.9 & 40.3 & 56.1 & 46.9 & \nan & \bfseries 0.0 & \nan & 83.1 & 61.3 & 70.5 & \nan & 0.0 & \nan \\
 & OWSM v3.1\m{+lang} & 63.3 & 71.8 & 24.1 & 35.1 & 28.6 & \bfseries 0.0 & \bfseries 0.0 & \bfseries 0.0 & 88.2 & 26.5 & 40.7 & \nan & 0.0 & \nan \\
 & \quad\m{+demucs} & 51.8 & 62.0 & 19.0 & 35.6 & 24.7 & \nan & \bfseries 0.0 & \nan & 83.7 & 27.5 & 41.4 & \nan & 0.0 & \nan \\
 \cmidrule{2-16}
 & \AudioShake\ v3 & 12.6 & 17.5 & 46.4 & 74.2 & 57.1 & 94.7 & 64.3 & 76.6 & 95.1 & 74.8 & 83.7 & 89.0 & 64.0 & 74.5 \\
 & JamendoLyrics & 5.0 & 37.6 & \nan & 0.0 & \nan & \nan & 0.0 & \nan & 98.7 & 95.8 & 97.2 & 95.9 & 85.4 & 90.3 \\
\midrule
\multirow[t]{12}{*}{French} & Whisper v2 & 27.7 & 31.1 & \bfseries 57.0 & 38.5 & \bfseries 45.9 & \nan & 0.0 & \nan & 89.5 & 62.2 & 73.4 & \bfseries 100.0 & \bfseries 0.1 & \bfseries 1.4 \\
 & \quad\m{+lang} & \bfseries 27.1 & \bfseries 30.5 & 55.7 & 38.2 & 45.3 & \nan & 0.0 & \nan & \bfseries 89.5 & 62.6 & 73.7 & \nan & 0.0 & \nan \\
 & \quad\m{+demucs} & 43.3 & 46.9 & 33.4 & \bfseries 44.4 & 38.0 & \nan & 0.0 & \nan & 83.5 & 54.5 & 66.0 & \nan & 0.0 & \nan \\
 & \qquad\m{+lang} & 38.2 & 42.1 & 30.9 & 43.5 & 36.1 & \nan & 0.0 & \nan & 84.2 & 53.8 & 65.6 & \nan & 0.0 & \nan \\
 & Whisper v3 & 34.7 & 38.0 & 56.5 & 34.1 & 42.5 & \nan & 0.0 & \nan & 78.3 & \bfseries 77.4 & \bfseries 77.9 & \nan & 0.0 & \nan \\
 & \quad\m{+lang} & 34.7 & 38.0 & 55.6 & 34.1 & 42.3 & \nan & 0.0 & \nan & 78.3 & \bfseries 77.4 & \bfseries 77.9 & \nan & 0.0 & \nan \\
 & \quad\m{+demucs} & 44.9 & 48.2 & 38.7 & 27.4 & 32.0 & \nan & 0.0 & \nan & 74.5 & 64.9 & 69.3 & \nan & 0.0 & \nan \\
 & \qquad\m{+lang} & 44.9 & 48.3 & 38.7 & 27.4 & 32.0 & \nan & 0.0 & \nan & 74.5 & 64.9 & 69.3 & \nan & 0.0 & \nan \\
 & OWSM v3.1\m{+lang} & 71.6 & 75.7 & 38.6 & 25.3 & 30.6 & \bfseries 10.0 & \bfseries 1.1 & \bfseries 1.9 & 77.4 & 23.4 & 36.0 & \nan & 0.0 & \nan \\
 & \quad\m{+demucs} & 78.5 & 82.1 & 22.2 & 22.5 & 22.3 & 0.0 & 0.0 & 0.0 & 86.0 & 26.8 & 40.9 & \nan & 0.0 & \nan \\
 \cmidrule{2-16}
 & \AudioShake\ v3 & 20.8 & 23.5 & 63.6 & 36.1 & 46.1 & 75.0 & 1.6 & 3.2 & 95.0 & 83.0 & 88.6 & 82.9 & 59.2 & 69.0 \\
 & JamendoLyrics & 10.3 & 23.3 & \nan & 0.0 & \nan & \nan & 0.0 & \nan & 98.4 & 91.3 & 94.7 & 91.4 & 93.9 & 92.6 \\
\bottomrule
\end{tabular}
\caption{Benchmark results (all metrics shown as percentages). WER is word error rate, \casewer\ is case-sensitive WER, the rest are precisions, recalls, and F-measures. ``\m{+demucs}'' indicates vocal separation using HTDemucs; ``\m{+lang}'' indicates that the language of each song was provided to the model instead of relying on auto-detection. Whisper results are averages over 5 runs with different random seeds, LyricWhiz over 2 runs; OWSM and \AudioShake\ are deterministic, hence the results are from a single run. The best results achieved by open-source systems are shown in {\bfseries bold}. LyricWhiz and \AudioShake\ are listed separately, because they rely on proprietary technology. The last row shows metrics computed between the original JamendoLyrics dataset and our revision.}
    \label{tab:full-results}
\end{table*}

\begin{table}
    \centering
    \scalebox{0.88}{%
    \renewcommand{\arraystretch}{0.9}
    \setlength\aboverulesep{0.35ex}
    \setlength\belowrulesep{0.65ex}
    \begin{tabular}{lr@{~}rr@{~}r@{~}rr@{~}r@{~}rr@{~}r@{~}r}
\toprule
& \multicolumn{2}{c}{Words} & \multicolumn{3}{c}{Punctuation} & \multicolumn{3}{c}{Line breaks} & \multicolumn{3}{c}{Section breaks} \\
\cmidrule(lr){2-3} \cmidrule(lr){4-6} \cmidrule(lr){7-9} \cmidrule(lr){10-12}
System & WER & \casewer & \ctr{$P_\punct$} & \ctr{$R_\punct$} & \ctr{$F_\punct$} & \ctr{$P_\nl$} & \ctr{$R_\nl$} & \ctr{$F_\nl$} & \ctr{$P_\sect$} & \ctr{$R_\sect$} & \ctr{$F_\sect$} \\
\midrule
Whisper v2 & \bfseries 34.5 & \bfseries 40.4 & \bfseries 36.4 & 51.3 & \bfseries 42.6 & \bfseries 95.5 & 50.7 & \bfseries 66.2 & --- & 0.0 & --- \\
\quad\m{+demucs} & 41.4 & 47.2 & 29.8 & \bfseries 52.4 & 38.0 & 89.5 & 46.8 & 61.4 & --- & 0.0 & --- \\
Whisper v3 & 59.0 & 63.8 & 35.2 & 46.3 & 40.0 & 75.3 & \bfseries 55.1 & 63.6 & --- & 0.0 & --- \\
\quad\m{+demucs} & 52.3 & 58.6 & 27.3 & 47.8 & 34.7 & 83.8 & 50.8 & 63.3 & 0.0 & 0.0 & 0.0 \\
OWSM v3.1 & 75.6 & 82.5 & 15.7 & 10.9 & 12.9 & 95.4 & 24.9 & 39.6 & \bfseries 100.0 & \bfseries 2.5 & \bfseries 4.9 \\
\quad\m{+demucs} & 82.9 & 91.8 & 20.2 & 14.7 & 17.0 & 82.1 & 25.8 & 39.2 & --- & 0.0 & --- \\
\midrule
\AudioShake\ v3 & 24.3 & 29.1 & 44.1 & 60.3 & 50.9 & 98.2 & 67.4 & 80.0 & 77.1 & 67.5 & 72.0 \\
    \bottomrule
    \end{tabular}
    }
    \caption{Full results on performance \texttt{SC06} from SWD.
    All systems are evaluated with the language (German) provided.
    Only punctuation (\texttt{P}), line breaks (\texttt{L}) and section breaks (\texttt{S}) are included, as the ground truth lyrics do not contain any parentheses.
    Whisper results are averages over 5 runs with different random seeds.
    The best results achieved by open-source systems (i.e.\ excluding \AudioShake) are shown in {\bfseries bold}.}
    \label{tab:full-swd-results}
\end{table}

\lstset{
    inputencoding=utf8,
    frame=single,
    basicstyle={\rmfamily\small},
    moredelim=[is][\underbar]{_}{_},
    columns=fullflexible,
    breaklines=true,
    breakatwhitespace=true,
    breakautoindent=true,
    breakindent=0em,
    showlines=true,
    literate=%
    {-}{-}1%
    {é}{{\'e}}{1}%
    {è}{{\`e}}{1}%
    {à}{{\`a}}{1}%
    {ç}{{\c{c}}}{1}%
    {œ}{{\oe}}{1}%
    {ù}{{\`u}}{1}%
    {É}{{\'E}}{1}%
    {È}{{\`E}}{1}%
    {À}{{\`A}}{1}%
    {Ç}{{\c{C}}}{1}%
    {Œ}{{\OE}}{1}%
    {Ê}{{\^E}}{1}%
    {ê}{{\^e}}{1}%
    {î}{{\^i}}{1}%
    {ô}{{\^o}}{1}%
    {û}{{\^u}}{1}%
    {ë}{{\¨{e}}}1
    {û}{{\^{u}}}1
    {â}{{\^{a}}}1
    {Â}{{\^{A}}}1
    {Î}{{\^{I}}}1
}

\begin{figure}
\centering
    \begin{minipage}[t]{0.51\linewidth}
\begin{lstlisting}
people gonna hate let them do it
shine like it ain't nothing to it
damn you a major influence
skate like there ain't nothing doing
live life don't say nothing to them

spectators
_side_ _liners_
spending days
coming up with sly comments
that's psychotic why _try a_ tarnish _such_ a fly product
why be mad just cause i got hey
i may never know
wave to the haters that put me on the pedestal talk smack
but they really know i'm incredible
unforgettable young blue eyes
the new guy is on schedule
man behind bars and _thats_ minus the federal
stone giant what the hell
could some pebbles do
while you revel in drama im building revenue
tell them you'll get them tomorrow _their_ ain't _nothing_ stressing you
life goes on _lifes_ goes on
you _was_ the shit even before those lights went on
they gonna trash you even if they like your song
people always gonna judge homie right or wrong
\end{lstlisting}
    \end{minipage}
    \hfill
    \begin{minipage}[t]{0.46\linewidth}
\begin{lstlisting}
People gon' hate, let 'em do it (_ah_)
Shine like it ain't nothin' to it (_that_'_s_ _right_)
Damn, you a major influence (_oh_)
Skate like there ain't nothin' doin'
Live life, don't say nothin' to 'em

Spectators, _sideliners_
Spendin' days comin' up with sly comments
That's psychotic, why tarnish a fly product?
Why be mad just 'cause I got it? Hey
I may never know, wave to the haters
That put me on the pedestal
Talk smack, but they really know I'm incredible
Unforgettable, young blue eyes, the new guy is on schedule
Man behind bars and _that_'_s_ minus the federal
Stone giant, what the hell could some pebbles do
While you revel in drama, I'm buildin' revenue
Tell 'em you'll get 'em tomorrow, _there_ ain't _no_ stressin' you
Life goes on, _life_ goes on
You the shit even before those lights went on
They gon' trash you even if they like your song
People always gon' judge homie right or wrong




\end{lstlisting}
    \end{minipage}
    \caption{An excerpt from \emph{Crowd Pleaser~-- Jason Miller} (license: CC BY-NC-SA). Left: JamendoLyrics, right: our revision. %
    Word edits (excluding letter case, formatting, punctuation and elisions) are underlined.
    }
    \label{fig:lyrics-ex-1}
\end{figure}

\begin{figure}
\centering
    \begin{minipage}[t]{0.51\linewidth}
\begin{lstlisting}
y'a pas que tes pas qui m'_inspire_
qui _roule_ qui se _cambre_ et se penchent
comme un danger qui m'attire

surtout t'_arrêtes_ pas tu sais que tout s'envolerait pour moi
t'es comme un soleil en été le monde tourne autour de toi
le jour la pluie les marais les saisons de chaud ou de froid
les guerres les paix les traités y'a le monde qui tourne et puis toi
y'a pas que tes pas qui m'_inspire_
belle j'ai vu des démons dans tes hanches
qui _roule_ qui se _cambre_ et se penchent
comme un danger qui m'attire








\end{lstlisting}
    \end{minipage}
    \hfill
    \begin{minipage}[t]{0.46\linewidth}
\begin{lstlisting}
Y a pas que tes pas qui m'_inspirent_
Qui _roulent_, qui se _cambrent_ et se penchent
Comme un danger qui m'attire

Surtout t'_arrête_ pas, tu sais
Que tout s'envolerait pour moi
T'es comme un soleil en été
Le monde tourne autour de toi
Le jour, la pluie, les marais
Les saisons de chaud ou de froid
Les guerres, les paix, les traités
Y a le monde qui tourne, et puis toi

Y a pas que tes pas qui m'_inspirent_
(_Y_ _a_ _pas_ _que_ _tes_ _pas_ _qui_ _m_'_inspirent_)
Belle, j'ai vu des démons dans tes hanches
(_Belle_, _j_'_ai_ _vu_ _des_ _démons_ _dans_ _tes_ _hanches_)
Qui _roulent_, qui se _cambrent_ et se penchent
(_Qui_ _roulent_, _qui_ _se_ _cambrent_ _et_ _se_ _penchent_)
Comme un danger qui m'attire
\end{lstlisting}
    \end{minipage}
    \caption{An excerpt from \emph{Pas que tes pas~-- AZUL} (license: CC BY-NC-SA). Left: JamendoLyrics, right: our revision. %
    Word edits (excluding letter case, formatting and punctuation) are underlined.}
    \label{fig:lyrics-ex-2}
\end{figure}

\fi

\end{document}